\documentclass[preprint2]{aastex}
\received{2003 June 27}
 \begin{document}

 \title{The Rotation Of The Deep Solar Layers}

 \author{S. Couvidat$^{1,2}$, R. A. Garc\'\i a$^2$, S. Turck-Chi\`eze$^2$, T.
 Corbard$^3$, C. J. Henney$^3$, \& 
 S. Jim\'enez-Reyes$^4$}

 \email{couvidat@stanford.edu, rgarcia@cea.fr, cturck@cea.fr}

 \affil{$^1$ W.W. Hansen Experimental Physics Laboratory, Stanford University,
 Stanford, CA 94305, USA\\
 $^2$ CEA/DSM/DAPNIA/SAp, CE Saclay, 91191 Gif sur Yvette, France\\
 $^3$ National Solar Observatory, 950 North Cherry Avenue, Tucson, AZ
 85726, USA\\
 $^4$ Themis, Instituto de Astrof\'\i sica de Canarias, c/V\'\i a L\'actea s/n,
 La Laguna, Tenerife, Spain}

 \shorttitle{Solar Rotation with the GOLF+MDI data}
 \shortauthors{Couvidat, Garci\'\i a, \& Turck-Chi\`eze}

 \begin{abstract}
From the analysis of low-order GOLF+MDI sectoral modes ($\ell \le 3, 6 \le n \le 15, |m|=\ell $) and LOWL data ($\ell > 3$), we derive the radial rotation profile assuming no latitudinal dependance in the solar core.
 These low-order 
 acoustic modes contain the most statistically significant information about rotation of the deepest solar layers and should be least influenced by internal variability associated with the solar dynamo. After correction of the sectoral splittings for their 
 contamination by the rotation of the higher latitudes, 
  we obtain a flat rotation profile down to 0.2 R$_{\odot}$.
 \end{abstract} 


 \section{Introduction}

 Helioseismologists use the oscillations of acoustic waves that propagate inside the Sun to infer its
 rotation profile.
 Due to the solar rotation (and magnetic fields), the frequencies of two modes of
 the same degree $\ell$ and radial 
 order $n$, but with different azimuthal orders $m$, are separated by a small
 amount referred to as splitting 
 ($\Delta \omega_{\ell n m}$).

 The understanding of the angular momentum redistribution in the deep interior
 requires a precise derivation 
 of the rotation profile below the convection zone, but this profile is still a
 matter of debate.
 In the past an increase of the rotation rate near the core has 
 long been favored (\mbox{e.g.} Lazrek et \mbox{al.} 1996), but recent results by
 Chaplin et \mbox{al.} (1999) and 
 Eff-Darwich, Korzennik, \& Jim\'enez-Reyes (2002) favor a slight decrease.
 Corbard et \mbox{al.} (1997), using LOWL data, and Ehgamberdiev et al. (2001),
 using IRIS data, 
 also derived such a decrease.

 Here we focus on the rotation in the solar core derived from the GOLF
 (Global Oscillations at Low 
 Frequency) and MDI (Michelson Doppler Imager) instruments, supplemented for modes
 with  $\ell >3$ by LOWL data. 
These data sets and the interest of the low-order modes are presented in sections 2 and 3 respectively. In section 4, the method utilized to extract the rotational splittings is discussed. In section 5 we discuss the rotation profile. We conclude in section 6.

 \section{The Data}

 We use data from GOLF/SoHO and MDI/SoHO for the modes $\ell \le 3$.  
 GOLF detects the global solar oscillations by the Doppler shift they produce at
 the surface on the sodium 
 lines. It is specifically designed to be more sensitive to the low-degree modes,
 and only detects those with $\ell+m$ equal to an even number.
 We utilize the 2034-day long GOLF series (starting in April 1996) for the GOLF-alone
 splittings. 
 MDI detects solar surface velocity from a nickel spectral line and resolves the
 solar surface. To produce 
 integrated-disk temporal series sensitive to the low-$\ell$ modes, we apply two masks
 (gaussian, and 
 gaussian zero mean) to the 2243-day long MDI series (Henney et
 \mbox{al.} 1999). These two  
 series have been properly combined with GOLF 2243-day long series to increase
 the signal-to-noise ratio of the 
 low-$\ell$ modes, and to reduce the leakage of higher-$\ell$ modes present on
 the MDI spectrum (Garc\'\i a et 
 \mbox{al.} 2003). The use of the MDI+GOLF time series allows for the precise measurement of lower-$n$
 modes. This point of interest is discussed further 
 in section 3.
 We supplemented these GOLF-alone and MDI+GOLF data with LOWL rotational splittings
 for $\ell > 3$. We used six 
 LOWL series of 1 year, from February 1994 to February 2000. The splittings have
 been 
 extracted for each year but the final result uses the 6 years statistics
 (Jim\'enez-Reyes 2001).

 \section{Information Value of Low-Order Modes}

\begin{figure}[ht]
 \plotone{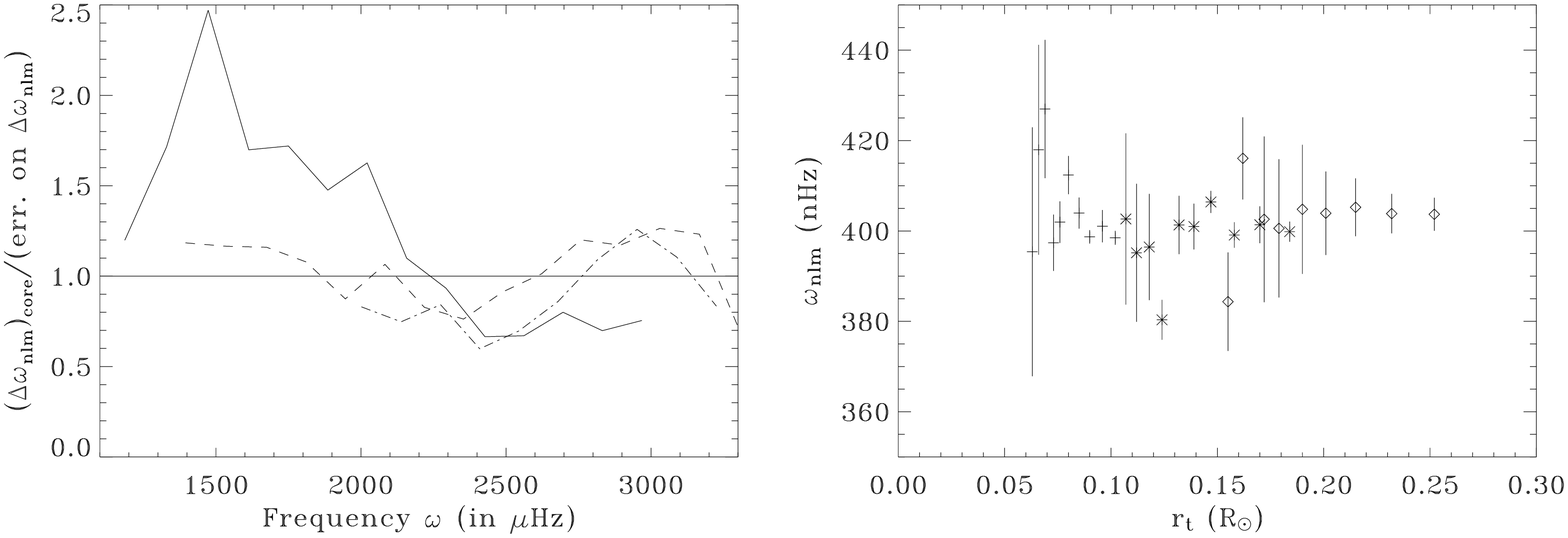}
 \epsscale{1.4}
 \caption{\label{fig1} Left panel: Ratio of the splitting due to the layers below
 $0.2 \, R_{\odot}$, to the uncertainty on the total splitting for $\ell=1$ (solid line), $\ell=2$ (dashed line), and $\ell=3$ (dash-dotted line).
 Right panel: synodic splittings for $n \le 15$ used for the inversion of the solar rotation,
 obtained with GOLF and MDI, as a function of the inner turning point for $\ell=1$ (crosses), $\ell=2$ (stars), and $\ell=3$ (diamonds).}
 \end{figure}

 The derivation of the rotation profile in the radiative zone requires
 high-precision splittings. For 
 instance, only $\simeq 3.5 \%$ of the splitting ($406$ nHz) of the mode $\ell=1$
 $n=9$ $m=1$ (at $1472.85 \, 
 \mu$Hz) is due to the rotation below $0.2 \, R_{\odot}$ (Couvidat 2002). Thus,
 we need to derive $\Delta 
 \omega_{1,9,1}$ with a precision better than $14.2$ nHz to obtain information
 on the core. 
 The error associated with this splitting is $\simeq \pm 5.3$ nHz (see Table
 \ref{tab1}). It contains more information 
 than the mode $\ell=1$ $n=20$ (at $2963.43 \, \mu$Hz), for which $6.5 \%$ of its
 splitting ($401$ nHz) is due to 
 the core, \mbox{i.e.} $26$ nHz, but the uncertainty is $34.6$ nHz. This point is
 illustrated by the left panel of Fig \ref{fig1}.
 Even though the inner turning point is closer to the core as $n$ increases, the
 outer turning point 
 is also closer to the surface, where stochastic excitation and solar cycle effects
 become more important.
 Therefore, the most favorable trade-off between sensitivity to the core rotation and
 uncertainty in the rotational splitting 
 determination occurs for low-$\ell$ low-$n$ p modes. Consequently we use
 only these modes in the present 
 analysis.

 \section{The Rotational Splittings}

 The splittings are obtained by fitting the components of each mode 
 to asymmetric Lorentzian profiles (Nigam \& Kosovichev 1998) using
  a maximum likelihood code based on Appourchaux, Gizon, \& Rabello-Soares (1998).
 The modes $\ell=0$ \& $2$ are fitted together, likewise the modes $\ell=1$ \& $3$. 
 For the GOLF-alone series, we use the periodogram of the data as the power spectrum, except for the modes $\ell=1$ $n=7$ \& $8$, and
 $\ell=2$ $n=8$ \& $9$, 
 for which we use a multitaper spectral estimator (\mbox{e.g.} Percival \& Walden 1993).
 For the combined MDI+GOLF series  
 we again use the multitaper.
 We restricted the search of the splittings to modes with a frequency less than
 $\simeq 3000 \, \mu$Hz, and use for the inversion only the modes with $n \le 15$. This 
 way we limit the impact of the solar cycle and surface effects. This approach is consistent with the results of
 Chaplin et \mbox{al.} (2001) who demonstrated 
 that the splittings are strongly biased at higher frequencies, and Chaplin et \mbox{al.} (2003), who showed that the solar cycle does not affect the splittings of the low-$n$ $\ell=2$ modes. 

 The free parameters of the fitted profiles for each mode are: the full width at
 half maximum (FWHM, the same 
 for all the components of a mode), the amplitude (the relative amplitudes of the
 different components are fixed 
 empirically), the asymmetry (the same for each component), the central
 frequency, the noise level (assumed to 
 be constant in the fit window of 40 $\mu$Hz), and the splittings. The latter parameters are sectoral values ($|m|=\ell$).
 We restrict the fits to frequencies $\le 2000-2200 \, \mu$Hz on the MDI+GOLF
 data, because of the growing influence of 
 higher-$\ell$ modes in the fit window (for instance, $\ell=5$ modes perturb the
 $\ell=2$ profiles).

 In Table \ref{tab1}, we list the synodic sectoral splittings and their
 uncertainties. The rotational splittings 
 common to the MDI+GOLF and GOLF-alone datasets generally agree within the
 $1\sigma$ uncertainty. 
 Right panel of Fig \ref{fig1} presents these splittings as a function of the
 inner turning point. This plot emphasizes the 
 high quality of the SoHO data after 5 years of observation in comparison with
 previous data sets.

 \begin{deluxetable}{llcclllccl}
 \tabletypesize{\scriptsize}
 \tablewidth{0pt}
 \tablecaption{\label{tab1} Synodic sectoral splittings ($\Delta \omega_{nlm}$)
 from GOLF-alone (left column) 
 and MDI+GOLF (right column). 
 The errors (err.) on the splittings are the $1\sigma$ uncertainties provided by
 the Hessian matrix of the fit. 
 $r_{t}$ is the inner turning point of the mode.}
 \tablehead{\colhead{$\ell$} & \colhead{$n$} & \colhead{$\Delta \omega_{nlm}$
 (nHz)} & \colhead{err. (nHz)} &  
 \colhead{$r_{t}$ ($R_{\odot}$)}& \colhead{$\ell$} & \colhead{$n$} &
 \colhead{$\Delta \omega_{nlm}$ (nHz)} & 
 \colhead{err.(nHz)} &  \colhead{$r_{t}$ ($R_{\odot}$)}}
 \startdata
 1 & 6 & ------/398.49 & ----/1.53 & 0.102 & 2 & 11& 401.07/401.33 & 8.20/6.45 & 0.132 \\
 1 & 7 & 393.39/401.05 & 6.46/3.57 & 0.096 & 2 & 12& 379.60/380.34 & 10.72/4.42& 0.124 \\
 1 & 8 & 400.67/398.70 & 6.06/1.49 & 0.090 & 2 & 13& 402.37/396.45 & 10.64/11.75& 0.118\\
 1 & 9 & 406.40/403.96 & 5.28/3.43 & 0.085 & 2 & 14& 395.17/------ & 15.27/-----& 0.112\\
 1 & 10& 414.63/412.37 & 9.50/4.22 & 0.080 & 2 & 15& 402.65/------ & 18.94/-----& 0.107\\
 1 & 11& 400.16/401.96 & 10.47/4.60& 0.076 & 3 & 6 & ------/403.69 & -----/3.64 & 0.252\\
 1 & 12& 406.56/397.39 & 14.19/6.22& 0.073 & 3 & 7 & ------/403.84 & -----/4.36 & 0.232\\
 1 & 13& 436.49/426.98 & 15.63/15.29&0.069 & 3 & 8 & ------/405.23 & -----/6.39 & 0.215\\
 1 & 14& 411.15/417.97 & 23.50/23.23&0.066 & 3 & 9 & ------/403.92 & -----/9.23 & 0.201\\
 1 & 15& 408.08/395.40 & 28.94/27.53&0.063 & 3 &10 & ------/404.79 & -----/14.28& 0.190\\
 2 & 6 & ------/399.84 & -----/2.23 & 0.184& 3 &11 & ------/400.58 & -----/15.30& 0.179\\
 2 & 7 & ------/401.37 & -----/4.07 & 0.170& 3 &12 & 401.52/402.58 & 10.10/18.34& 0.172\\
 2 & 8 & 397.00/399.10 & 5.70/2.83 & 0.158 & 3 &13 & 388.53/416.07 & 11.47/9.09 & 0.162\\
 2 & 9 & 412.01/406.42 & 6.49/2.43 & 0.147 & 3 &14 & 413.62/384.35 & 11.65/10.90& 0.155\\
 2 & 10& 406.55/400.97 & 6.95/5.06 & 0.139 & 3 &15 & 409.72/410.05 & 16.37/20.18& 0.148\\

 \enddata
 \end{deluxetable}

 \section{The Solar Rotation Profile}

\begin{figure}[!ht]
 \centering
 \epsscale{0.95}
 \plotone{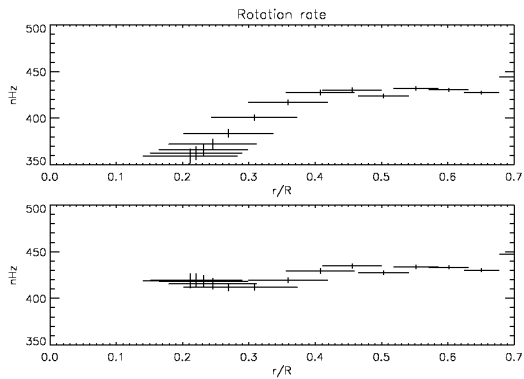}
 \caption{\label{fig2} Upper panel: rotation profile directly derived with GOLF \&
 MDI, combined with  LOWL 
 splittings for $\ell > 3$.
 Lower panel: rotation profile obtained after correction of the splittings to
 account for the differential 
 rotation in the convective zone.
 The vertical error bars result from propagating splitting measurement errors
 through the inversion process, while 
 the horizontal error bars give the FWHMs of the corresponding averaging kernels,
 and is an estimate of the 
 resolution achieved at each depth.}
 \end{figure}

 The 1D inversion of sectoral splittings provides information about the rotation mainly
 along the solar equator, but with contamination by the higher latitudes. We applied a 1D MOLA inversion
 technique (Multiplicative Optimally 
 Localized Averages, see \mbox{e.g.} Corbard et \mbox{al.} 1998). 
 The MDI+GOLF and GOLF-alone splittings (when the former are unavailable) were 
 completed for $\ell > 3$ with splittings from LOWL.
 The direct inversion of the data leads to a rigid rotation from the base of the
 convective zone at $0.713 \, 
 R_{\odot}$ down to $\simeq 0.35 \, R_{\odot}$ (see upper panel of Fig
 \ref{fig2}). Below there is a decrease 
 in the rotation rate of the solar equator, as mentioned previously by several authors.
 Depending on the regularization 
 parameter, this decrease is more or less pronounced (here we show the profile with the most pronounced decrease). 
 However, the 1D inversion with the sectoral splittings relies on the assumption
 that the rotation rate at any latitude is equal to the rate at the equator. 

As the differential rotation of the convective zone implies the higher latitudes
 rotate more slowly than the equator, we 
 need to correct the splittings for this contribution. 
 Therefore, we add $12$ nHz to the $\ell=1$ splittings, $8$ nHz to the $\ell=2$, and
 $6$ nHz to the 
 $\ell=3$, following Corbard et \mbox{al.} (1998). The values of these corrections were derived from a 2D Regularized Least-Squares inversion of MDI data in the convective zone, by comparing the splittings computed with the exact linear relation (that involves the 2D rotational kernels and rotation profile) with the splittings computed with the 1D sectoral approximation.
 We present on the lower panel of Fig \ref{fig2} the profile derived with the
 corrected splittings. We obtain a flat
 profile in the radiative interior. The dip around $0.3 \, R_{\odot}$ might be
 due to the absence of correction 
 of the LOWL splittings but we cannot rule out a physical phenomenon at the limit of the nuclear core.
 
This careful inversion of our most precise seismic data therefore favors a flat rotation
 curve at $430$ nHz in the solar 
 radiative interior. 
 This lends support for a magnetic field strong enough to suppress any differential rotation
 that might arise from an angular momentum redistribution through the gravity waves (Talon, Kumar, \& Zahn 2002).
 We think that the absence of a decrease or increase in the rotation rate, compared with previous works, is mainly
 due to the quality of the present data and the limitation in the degradation of
 these data by the external layers.
 We note that an increase of the rotation rate seems now very unlikely down to
 $0.2 \, R_{\odot}$, 
 but cannot be ruled out below $0.2$ by the present gravity-mode analysis (Turck-Chi\`eze et
 \mbox{al.} 2002).

We have also reduced the MDI+GOLF splittings to $a_1$ coefficients (Schou 
et \mbox{al.} 1994)
by removing $a_3$ (for $\ell=2,3$) and $a_5$ (for $\ell=3$) estimated from our knowledge 
of the rotation in the convection zone.
This allows us to combine them directly with LOWL $a_1$ coefficients and 
carry a 1D inversion
for a latitudinally averaged rotation rate. This gives results nearly identical to the one shown on the bottom panel of Fig \ref{fig2}.
Finally, we also suppressed the solar cycle effects for intermediate- and high-$\ell$ by using only the first year of LOWL data. Of course this increases the vertical error bars but does not change the rotation profile.

 \section{Summary}

 By using the high-quality seismic data from the spatial instruments GOLF \&
 MDI, and by using only the low-$n$ modes, 
 we limit the effects of the variable magnetic field that takes place in the
 outer layers and obtain a very coherent dataset for rotation below the solar convection zone. While a proper treatment 
 of the latitudinal dependence of the rotation speed is still needed to extract the
 rotation in the core from 1D inversions, the method we apply here removes the effect of latitudinal variation of the rotation in the convection zone, and we obtain a flat rotation profile down to 0.2
 $R_{\odot}$. This puts a strong 
 constraint on the redistribution of the angular momentum. The
 uncertainties in the rotation rate are still quite large 
 below $0.3 \, R_{\odot}$: progress on this point can be achieved by the
 detection of mixed low-$\ell$ pressure modes, and gravity modes.

 \section*{Acknowledgments}
 GOLF \& MDI instruments are cooperative efforts (of French and Spanish
 institutes for GOLF) to whom we are deeply 
 indebted. SoHO is a project of international cooperation between ESA and NASA.
 S.C. acknowledges the support of 
 a CEA/Saclay doctoral Grant and the NASA Grant NAG5-10483. We wish to thank ECHO
 team for providing us the LOWL 
 mode parameters.
 

 \end{document}